\documentclass[conference,letterpaper]{IEEEtran}
\IEEEoverridecommandlockouts
\usepackage[T1]{fontenc}
\usepackage[scaled=0.96]{helvet}
\usepackage{graphicx,amsmath,booktabs,array,tabularx,url,cite}
\usepackage[hidelinks]{hyperref}
\hypersetup{pdftitle={KPG193-ext: A Synthetic Korean Power Grid Test System for Stability Analysis},
pdfauthor={Sungyoon Song, Geonho Song, Min-han Yoon, Yong-Hyun Song, Seung Wan Kim, Jip Kim},
pdfsubject={PESGM 2027 manuscript v1.2; nuclear governor K=0; chronological stability assessment}}

\graphicspath{{figures/}}
\newcommand{\pv}{PV--wind}
\newcommand{\KPG}{\texttt{KPG193-ext}}
\newcommand{\KPGbase}{\texttt{KPG193}}
\newcommand{\paperTableStyle}{\sffamily\fontsize{8}{10}\selectfont\renewcommand{\arraystretch}{1.12}\setlength{\tabcolsep}{5pt}\setlength{\heavyrulewidth}{0.6pt}\setlength{\lightrulewidth}{0.35pt}}
\newcommand{\opcase}[1]{\textsf{#1}}

\makeatletter
\renewcommand{\@IEEEfigurecaptionsepspace}{\vskip4pt\relax}
\let\kpg@originalmakecaption\@makecaption
\long\def\@makecaption#1#2{%
  \ifx\@captype\@IEEEtablestring
    \footnotesize\bgroup\par\centering
    \@IEEEtabletopskipstrut{\normalfont\footnotesize #1.\nobreakspace\scshape #2}\par
    \addvspace{0.5\baselineskip}\egroup\@IEEEtablecaptionsepspace
  \else\kpg@originalmakecaption{#1}{#2}\fi}
\makeatother

\begin{document}

\title{\KPG{}: A Synthetic Korean Power Grid\\Test System for Stability Analysis
\thanks{This work was supported by NEXT group. *Corresponding author: Jip Kim (e-mail: \href{mailto:jipkim@kentech.ac.kr}{jipkim@kentech.ac.kr}).}}

\author{\IEEEauthorblockN{Sungyoon Song\textsuperscript{1}, Geonho Song\textsuperscript{2}, Min-han Yoon\textsuperscript{3}, Yong-Hyun Song\textsuperscript{4}, Seung Wan Kim\textsuperscript{4}, Jip Kim\textsuperscript{2,*}}
\IEEEauthorblockA{\textsuperscript{1}\textit{Tech University of Korea (TU Korea), Siheung, Korea}\\
\textsuperscript{2}\textit{AGM Center, Korea Institute of Energy Technology (KENTECH), Naju, Korea}\\
\textsuperscript{3}\textit{Kwangwoon University, Seoul, Korea}\\
\textsuperscript{4}\textit{NEXT group, Seoul, Korea}}}

\maketitle
\pagestyle{plain}\thispagestyle{plain}
% Keep normal interparagraph spacing instead of stretching short columns.
\raggedbottom

\begin{abstract}
Growing shares of inverter-based generation and constrained transmission expansion increase the need to assess stability across operating conditions. Production-cost models provide chronological commitment and dispatch, but stability analysis requires corresponding electrical detail and device dynamics. This paper introduces \KPG{}, an extension of the synthetic Korean power-grid test system \KPGbase{} that links chronological operating points with detailed electrical and dynamic models. The extension adds synchronous-machine, controller and renewable-converter models, together with explicit voltage levels, generation terminals, inter-voltage and step-up transformers, and reactive compensation. The resulting 731-bus network retains the original 193 regional locations. Daily unit commitment followed by hourly AC optimal power flow establishes 8,760 operating points for 2025, each distributed as paired steady-state and dynamic-model files. Selected-hour studies examine generation-loss frequency response, steady-state flow--voltage characteristics, and rotor-angle response with special protection system actions. The results reveal differences in delivered frequency support, internal voltage stress along regional transfer paths, and the angle--frequency tradeoff of corrective generator tripping. The database provides a common basis for examining the stability implications of commitment, dispatch and corrective action across daily and seasonal operating conditions.
\end{abstract}


\section{Introduction}

Power-system planning and operation seek to meet demand at least cost while satisfying reliability requirements. Production-cost models optimize commitment and dispatch under operating constraints; stability analyses examine how the system responds to disturbances at the resulting operating points. These assessments are connected through the generation schedule, which determines the machines in service, their loading and network transfers.

Hourly changes in demand and renewable output alter the operating conditions under which stability must be assessed. Recent planning studies and guidelines follow these changes by translating production-cost schedules into detailed AC network cases~\cite{doe2024ntps,epri2024pcm_pf}. Extending this process to stability analysis requires a common data basis linking each operating point with its electrical equipment and dynamic models.

Synthetic-grid methods address these requirements through reactive-power planning and dynamic-model assignment~\cite{birchfield2018reactive,xu2018dynamics}. For Korea, \KPGbase{} provides a regional network for chronological unit commitment (UC) and AC optimal power flow (AC-OPF)~\cite{song2025kpg193}. Extending it to stability assessment requires the transformers and reactive compensation omitted from the original model, together with machine and controller dynamics. These additions must remain associated with the hourly generation schedules.

This paper introduces \KPG{} with three contributions: (i) synchronous-machine, control and converter models with documented formulations and synthetic parameter assignments; (ii) a 731-bus electrical representation with explicit voltage levels, generation terminals, transformers and reactive compensation, preserving the original 193 regional locations; and (iii) 8,760 hourly operating points for 2025, obtained from daily UC and hourly AC-OPF and distributed as paired steady-state and dynamic-model files. Illustrative frequency, steady-state flow--voltage and rotor-angle studies demonstrate how this database connects scheduling decisions with stability assessment across operating conditions.

\begin{figure*}[t]
\centering\includegraphics[width=\textwidth]{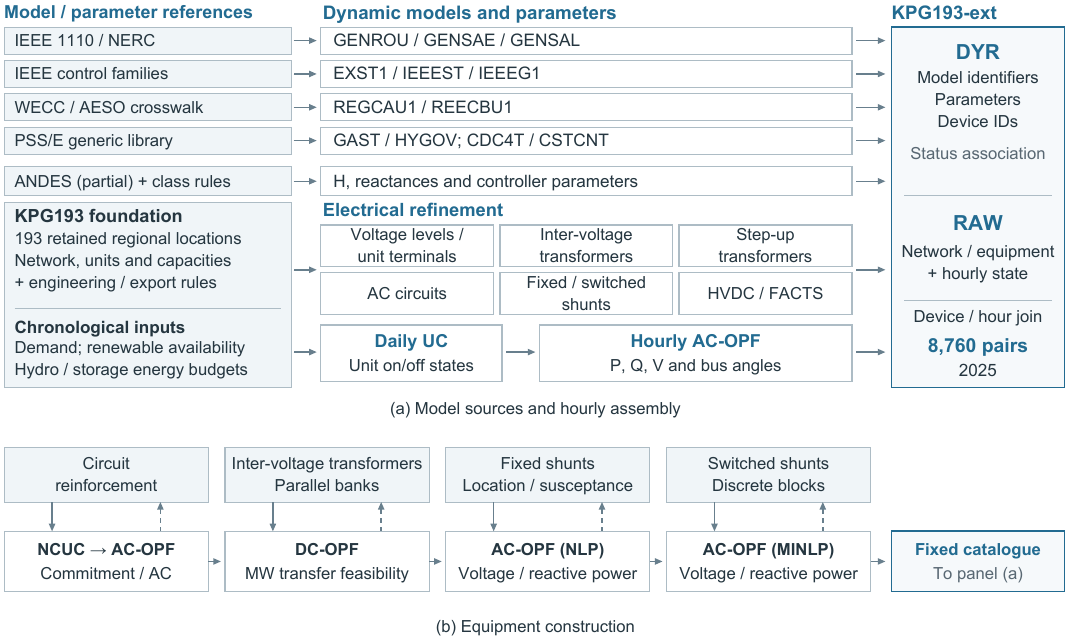}
\caption{Construction of \KPG{}. (a) Model and parameter sources, electrical refinement, and hourly RAW--DYR assembly. (b) Equipment-design iterations: DC-OPF screens MW transfer through candidate inter-voltage transformer banks; AC-OPF assesses voltage, reactive power and discrete shunt blocks. The resulting equipment catalogue supplies panel (a). Solid arrows pass configurations through the checks; dashed arrows denote equipment revisions. NCUC denotes network-constrained unit commitment.}
\label{fig:construction}
\end{figure*}

\section{Construction of \KPG{}}

Extending \KPGbase{} to stability analysis requires detailed network equipment and device dynamics consistent with its chronological operating points. The 2025 \KPGbase{} release supplies the regional topology, generation fleet, demand profiles and renewable availability used as the common basis for this extension~\cite{song2025kpg193}. Figure~\ref{fig:construction}(a) summarizes how these inputs are supplemented and assembled into hourly cases.

The network is first refined by separating transmission voltage levels and generation terminals within each regional location. Inter-voltage and step-up transformers connect these buses, and lower-voltage circuits omitted by the original regional representation are restored along existing corridors. Fixed and switched shunts and FACTS compensators are added to represent reactive support. These additions preserve the geographic aggregation while providing the electrical connections and voltage-control equipment needed for stability studies. Table~\ref{tab:models} gives the resulting inventory.

Equipment selection is iterated with the scheduling and network calculations in Fig.~\ref{fig:construction}(b). Network-constrained UC supplies schedules to AC-OPF, with circuit reinforcement revisited as needed. After voltage-level separation, DC-OPF screens transfers through candidate transformer banks; adjusting the number of parallel banks changes both capacity and equivalent reactance. Subsequent AC-OPF stages address voltage and reactive-power requirements through fixed-shunt placement and sizing, followed by discrete switched-shunt selection. These iterations reconcile scheduled transfers with the refined electrical network and establish the equipment catalogue used for hourly assembly.

Dynamic models are then assigned by generation and device class using IEEE-related machine and controller families, WECC renewable-converter formulations, and the applicable PSS/E library models~\cite{nercModels,psseModels,aesoModels}. Their numerical parameters draw on the ANDES reference where applicable, supplemented by generation-class and device assumptions~\cite{andes190modelreference}. These formulations and synthetic parameter assignments provide dynamic models for electromechanical stability studies. For each synchronous generator, the source reactance in the RAW case matches the subtransient reactance of the assigned dynamic model on the machine MVA base; the step-up transformer is represented separately.

Daily UC and hourly AC-OPF establish 8,760 operating points for 2025, each distributed as a RAW steady-state case paired with a DYR dynamic-model file. Consistent device identifiers and machine bases associate the two, and the AC operating point supplies terminal conditions for dynamic initialization. 
The paired files connect production-cost scheduling with stability studies at each hour. 
AC power-flow re-solves confirmed numerical convergence for all 8,760 hourly snapshots. These paired cases support comparisons across hourly schedules and assessments of redispatch or corrective action starting from a common operating point.
The dataset is available from the AGM Center~\cite{kpg193extDownload}. 

\begin{table}[t]
\caption{Model and Equipment Inventory}
\label{tab:models}\centering\paperTableStyle
\begin{tabularx}{\columnwidth}{@{}>{\raggedright\arraybackslash}p{.28\columnwidth}>{\raggedright\arraybackslash}X@{}}
\toprule Group & Representation\textsuperscript{*} \\
\midrule
Conventional & GENROU; EXST1; GAST (LNG) or IEEEG1 (coal/nuclear); IEEEST \\
Hydro / storage & GENSAE / GENSAL; EXST1; HYGOV \\
Aggregate \pv & REGCAU1; REECBU1; VTGTPAT or FRQTPAT\textsuperscript{\textdagger} \\
HVDC / FACTS & CDC4T / CSTCNT \\
\midrule
Electrical buses & 731: 293 transmission; 245 unit; 193 renewable \\
AC circuits & 820: 746 retained; 74 restored \\
Transformers & 240 inter-voltage; 438 step-up \\
Reactive shunts & Fixed and switched compensation \\
\bottomrule
\end{tabularx}
\vspace{3pt}\par\raggedright\sffamily\fontsize{7}{8.5}\selectfont
\textsuperscript{*}Library model identifiers with synthetic parameter assignments; formulation and parameter provenance are distinguished in Sec.~II. Counts refer to the electrical catalogue. The 74 restored circuits comprise 72 at 154 kV and two at 345 kV.\par\textsuperscript{\textdagger}Optional protection models; not enabled in the reported simulations.
\end{table}

\begin{figure*}[t]
\centering\includegraphics[width=\textwidth]{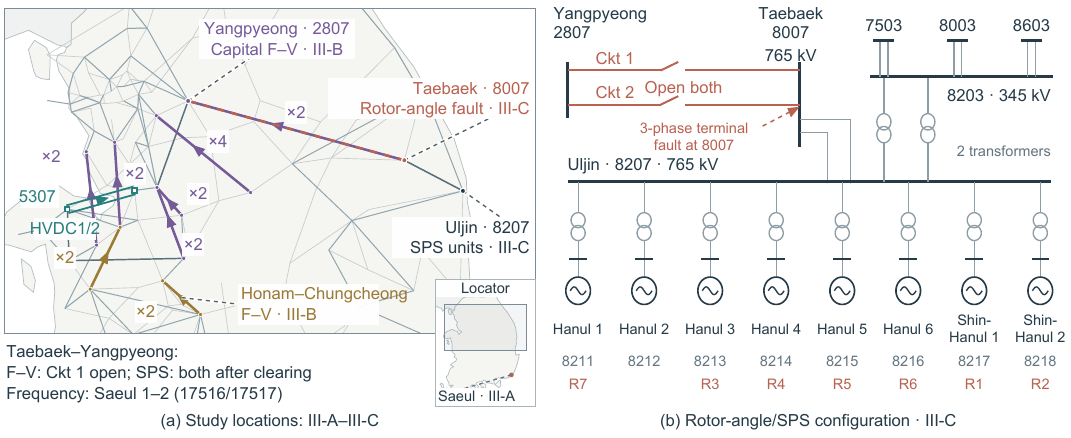}
\caption{Study locations and SPS configuration. (a) Saeul generation-loss events (III-A; locator), regional F--V interfaces (III-B), and the Taebaek rotor-angle fault (III-C). Capital AC circuits are purple, Honam--Chungcheong circuits ochre, and HVDC links teal. Arrows indicate transfer directions; $\times n$ denotes circuit counts. On Taebaek--Yangpyeong, III-B opens circuit~1 and III-C opens both circuits after fault clearing. (b) Taebaek 8007 terminal fault and generator connections; both Taebaek--Uljin 8207 circuits remain connected. R1--R7 specify the tested SPS order, skipping offline units; selected generators trip simultaneously. Transmission circuits are drawn individually in (b). Locations are regional and paths schematic.}
\label{fig:map}
\end{figure*}

\section{Illustrative Stability Studies}

An hourly generation schedule determines the committed fleet and available corrective actions, while its spatial distribution shapes regional transfers. The following studies use the electrical and dynamic representations of \KPG{} to examine three consequences of that schedule: frequency support after generation loss, voltage constraints along transfer paths, and the angle and frequency effects of special protection system (SPS) generator tripping.

The operating hours in Table~\ref{tab:cases} provide complementary conditions. Spring daytime \opcase{S} captures high renewable output, and \opcase{I} provides a low-synchronous-energy comparison with both frequency-study units online. \opcase{P} supplies a high-demand reference and the capital-area voltage-study operating point. Spring nighttime \opcase{T} adds the maximum Taebaek--Yangpyeong export among spring hours for the SPS comparison.

\begin{table}[t]
\caption{Selected Hourly Operating Conditions in 2025}
\label{tab:cases}\centering\paperTableStyle
\begin{tabularx}{\columnwidth}{@{}ll>{\raggedright\arraybackslash}X@{}}
\toprule Case & Date / hour & Selection criterion \\
\midrule
\opcase{P} & Aug 25 15:00 & Maximum system demand (100.68~GW) \\
\opcase{S} & Apr 26 13:00 & Maximum dispatched \pv\ share (44.43\%) \\
\opcase{I} & May 25 13:00 & Minimum synchronous energy (209.70~GW$\cdot$s)\textsuperscript{\textdagger} \\
\opcase{T} & Mar 16 22:00 & Spring maximum Taebaek--Yangpyeong export \\
\bottomrule
\end{tabularx}
\vspace{3pt}\par\raggedright\sffamily\fontsize{7}{8.5}\selectfont
% \textsuperscript{*}All dates are in 2025. Hour labels follow case filenames; timezone and interval interpretation remain unconfirmed. Selection uses the released hourly snapshots; each selected case is re-solved before simulation.
\textsuperscript{\textdagger}Minimum among hours with both Saeul Units 1--2 committed.
% ; dispatched \pv\ share breaks ties.
\end{table}

Figure~\ref{fig:map}(a) identifies the generation-loss site, regional interfaces and fault corridor; Fig.~\ref{fig:map}(b) shows the individual generator connections and tested SPS order. Generation-loss studies use seasonal ZIP loads, while rotor-angle studies use constant-current active and constant-impedance reactive demand. Nuclear units retain their dispatch and inertia but provide no governor response (IEEEG1 $K=0$). Both enable frequency dependence of network parameters and use matching no-event checks.

\begin{figure*}[t]
\centering\includegraphics[width=\textwidth]{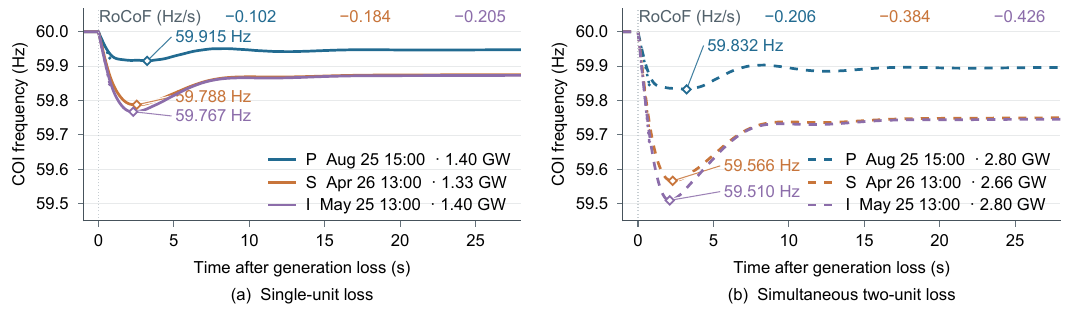}
\caption{Center-of-inertia frequency after (a) Saeul Unit~1 loss (bus 17516) and (b) simultaneous loss of Units~1--2 (buses 17516/17517). Dated legends identify operating conditions and the pre-event dispatch removed. The initial rate of change of frequency (RoCoF) is fitted over the first 100~ms; dotted lines extend the fitted slopes to 0.75~s. Hollow diamonds mark the sampled nadirs. Both panels use identical condition colors and frequency scales.}
\label{fig:frequency}
\end{figure*}
\begin{figure*}[t]
\centering\includegraphics[width=\textwidth]{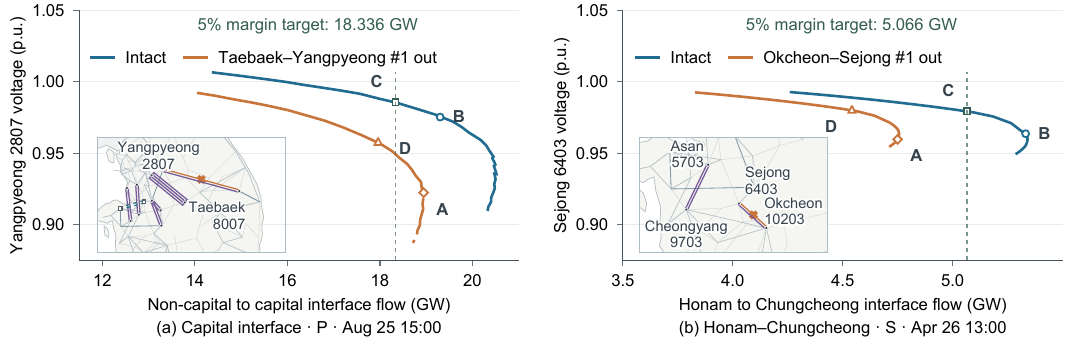}
\caption{F--V characteristics for (a) capital-area imports in \opcase{P} (Aug. 25, 15:00, 2025): 14 AC circuits and two HVDC links, with Taebaek 8007--Yangpyeong 2807 circuit~1 out; and (b) Honam--Chungcheong transfers in \opcase{S} (Apr. 26, 13:00, 2025): Cheongyang--Asan and Okcheon--Sejong, two circuits each, with Okcheon 10203--Sejong 6403 circuit~1 out. Receiving-terminal voltages are monitored at Yangpyeong 2807 and Sejong 6403, respectively. Interface flow is measured at receiving terminals, including the HVDC receiving-converter AC terminals. Insets show neighboring routes (pale), studied AC circuits (purple), HVDC links (dashed), and the opened circuit (orange cross). A is a solved contingency reference in (a) and a reactive-limit-associated turning state in (b). B is intact at the command of A; C has 95\% of the flow at B; D is the outage solution at the command of C.}
\label{fig:voltage}
\end{figure*}

\subsection{Generation-Loss Frequency Response}

The frequency study examines the support delivered by the scheduled fleet after generation loss. Loss of Saeul Unit~1 [Fig.~\ref{fig:frequency}(a)] and simultaneous loss of Units~1--2 [Fig.~\ref{fig:frequency}(b)] represent 1.4- and 2.8-GW-class contingencies. Center-of-inertia (COI) frequency is the kinetic-energy-weighted mean frequency of the surviving synchronous generators.

For the same 1.40-GW single-unit loss, the initial decline in \opcase{I} is approximately twice as steep as in \opcase{P}, consistent with its smaller synchronous energy. The nadir also reflects the subsequent mechanical-power and load response. \opcase{S} loses less generation than \opcase{P} but reaches a lower nadir: one second after the event, mechanical power has increased by 0.65~GW in \opcase{S}, compared with 1.03~GW in \opcase{P}. The dynamic models reveal differences in delivered support that the scheduled outputs alone do not specify.

The two-unit events deepen the nadirs while retaining the ordering \opcase{I}, \opcase{S} and \opcase{P} from lowest to highest. The comparison thus connects each hourly schedule to both the initial frequency decline and the subsequent response of the surviving fleet.

\begin{figure*}[t]
\centering\includegraphics[width=\textwidth]{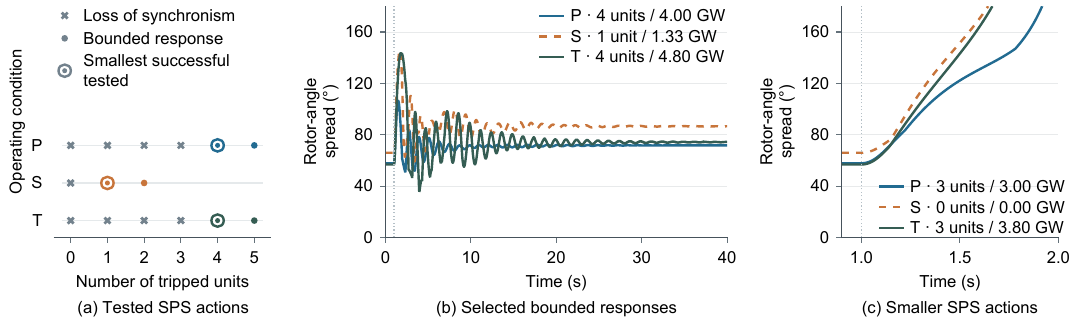}
\caption{SPS responses to the Taebaek--Yangpyeong two-circuit outage in \opcase{P} (Aug. 25, 15:00), \opcase{S} (Apr. 26, 13:00) and \opcase{T} (Mar. 16, 22:00), 2025. (a) Tested actions following the order in Fig.~\ref{fig:map}(b): crosses denote loss of synchronism assessed by the out-of-step monitor in Sec.~III-C; dots, bounded 40-s responses; rings, smallest successful tested actions. (b) Selected bounded responses. (c) Responses with one fewer SPS-initiated generator trip. Panels (b)--(c) show maximum--minimum rotor-angle spread among surviving synchronous generators; classification uses individual COI-relative angles.}
\label{fig:angle}
\end{figure*}

\subsection{Steady-State Flow--Voltage Assessment}

The flow--voltage (F--V) studies examine how regional transfers interact with reactive support in the refined network. Capital-area imports in \opcase{P} [Fig.~\ref{fig:voltage}(a)] and Honam--Chungcheong transfers in \opcase{S} [Fig.~\ref{fig:voltage}(b)] contrast concentrated demand with spring renewable export.

Conventional generation is increased outside the capital area and reduced inside for the capital study, and increased in Honam and reduced in the northern regions for the spring study. Demand, commitment, renewable output and HVDC schedules remain fixed. Reactive-power limits and STATCOM/HVDC controls remain active, while generator MW bounds are relaxed for tracing.

The A--D correspondence in Fig.~\ref{fig:voltage} applies the 5\% margin to the intact-system flow at B, giving $F_{\mathrm{target}}=0.95F_B$. Intact C and outage D are independently solved at the same redispatch command, preserving the operating-point correspondence despite differences in flows and losses.

The mapped targets are 18.336~GW for the capital interface and 5.066~GW for Honam--Chungcheong, compared with initial flows of 14.367 and 4.259~GW. These are path-specific F--V targets: the capital reference is not a confirmed loadability limit.
At A, near the spring turning region, an aggregate renewable generator reaches its reactive-power maximum while receiving-terminal Sejong remains at 0.960~p.u.; the interface-terminal voltage alone does not indicate this internal reactive-support limit.

\subsection{Rotor-Angle Response and SPS Generator Tripping}

Fast generator tripping via SPS can preserve synchronism after an export-corridor fault, but the resulting generation deficit also affects frequency. The Taebaek--Yangpyeong study evaluates this tradeoff across hourly operating conditions. A three-phase fault at Taebaek at 1~s is cleared after five cycles by opening both circuits; selected Uljin units trip at 1.15~s. The disturbance and SPS order in Fig.~\ref{fig:map}(b) are fixed across \opcase{P}, \opcase{S} and \opcase{T}, with offline units skipped.

Figure~\ref{fig:angle} plots the raw rotor-angle spread, $S_{\delta}(t)=\max_i\delta_i(t)-\min_i\delta_i(t)$, among synchronous generators retained after SPS. Classification uses a separate COI-relative monitor adapted from the simplified relative-angle approach in~\cite{nerc2018irol}: a retained machine reaches $\lvert\delta_i-\delta_{\mathrm{COI}}\rvert\geq120^{\circ}$ while its outward relative frequency exceeds 0.01~Hz. The 120$^{\circ}$ threshold applies to individual COI-relative angles, not to the maximum--minimum spread plotted in Fig.~\ref{fig:angle}. Bounded responses require no relative pole slips and decaying oscillations over 40~s.

The smallest successful tested actions in Fig.~\ref{fig:angle}(a) trip four units in \opcase{P}, one in \opcase{S} and four in \opcase{T}. Although initial corridor flows in \opcase{P} and \opcase{S} are similar (5.470 and 5.159~GW), their successful actions differ markedly. This contrast illustrates the need to evaluate SPS actions against the full hourly operating point, including unit availability and dispatch.

The selected actions arrest angular separation and produce decaying oscillations [Fig.~\ref{fig:angle}(b)]. In \opcase{T}, the peak-to-peak spread variation decreases from about 10$^{\circ}$ over 15--20~s to below 1$^{\circ}$ over 35--40~s. 
With one fewer SPS-initiated generator trip, angular separation instead grows and triggers the out-of-step monitor in all three conditions [Fig.~\ref{fig:angle}(c)].

Table~\ref{tab:actions} shows that the same four-unit trip count in \opcase{P} and \opcase{T} entails different generation losses and frequency nadirs.
Within \opcase{T}, adding a fifth trip reduces the maximum angle spread from 143$^{\circ}$ to 110$^{\circ}$ but deepens the nadir from 59.28 to 59.20~Hz. 
Additional tripping therefore improves the angle response at a frequency cost. By linking individual unit connections and dynamic models with hourly commitment and dispatch, \KPG{} enables the corrective action and its frequency consequence to be assessed together.

\begin{table}[t]
\caption{Rotor-Angle Spread and COI Frequency After SPS}
\label{tab:actions}\centering\paperTableStyle
\setlength{\tabcolsep}{2pt}
\begin{tabularx}{\columnwidth}{@{}lXrrr@{}}
\toprule
Case / hour & SPS set\textsuperscript{*} & Loss (GW) & $S_{\delta,\max}$ ($^{\circ}$) & Nadir (Hz) \\
\midrule
\opcase{P} / Aug 25 15:00 & H3--6  & 4.00 & 106 & 59.73 \\
\opcase{S} / Apr 26 13:00 & SH1  & 1.33 & 146 & 59.70 \\
\opcase{T} / Mar 16 22:00 & \mbox{H3--4,} \mbox{SH1--2}  & 4.80 & 143 & 59.28 \\
\bottomrule
\end{tabularx}
\vspace{3pt}\par\raggedright\sffamily\fontsize{7}{8.5}\selectfont
\textsuperscript{*}H/SH denote Hanul/Shin-Hanul; dates are in 2025. Loss is the pre-event dispatch of SPS-tripped units. $S_{\delta,\max}$ is the peak maximum--minimum rotor-angle spread among surviving synchronous generators over 1--30~s. Nadir is the minimum surviving-machine COI frequency over 1.15--30~s.
\end{table}

\section{Conclusion}
\KPG{} extends \KPGbase{} with device dynamics and detailed electrical equipment, providing 8,760 paired steady-state and dynamic-model cases for 2025. The illustrative studies demonstrate the value of keeping these representations associated with hourly commitment and dispatch. Frequency support and successful SPS actions vary across operating conditions, while additional generator tripping can improve the angle response at a frequency cost. Steady-state F--V analysis also reveals internal voltage stress beyond the monitored receiving terminals. The database thus provides a common basis for examining both the stability consequences of generation schedules and the tradeoffs of corrective action.

\bibliographystyle{IEEEtran}
\bibliography{references}
\end{document}